\documentclass[11pt]{article}
\usepackage{a4wide}       
\usepackage{amsmath,amssymb,amsfonts}
\usepackage{comment}
\usepackage{epsfig,verbatim,graphics}
\usepackage{slashed}

\newcommand{\mathsym}[1]{{}}

\newcommand{\eref}[1]{(\ref{#1})}

\renewcommand\({\left(}
\renewcommand\){\right)}
\renewcommand\[{\left[}
\renewcommand\]{\right]}
\newcommand{\pa}{\partial}
\newcommand{\dd}{{\rm d}}

\newcommand\mpl{m_{\rm p}}

\def\ba{\begin{eqnarray}}
\def\ea{\end{eqnarray}}
\def\be{\begin{equation}}
\def\ee{\end{equation}}

\def\M{\mathcal{M}}

\def\nablamu{\nabla^\mu}
\def\nablanu{\nabla^\nu}
\def\nn{\nonumber}
\def\({\left(}
\def\){\right)}

\def\eref#1{(\ref{#1})}

\newcommand{\roughly}[1]{\mathrel{\raise.3ex\hbox{$#1$\kern-0.85em
\lower1ex\hbox{$\sim$}}}}
\newcommand{\bphi}{\bar{\phi}}
\newcommand{\bPhi}{\bar{\Phi}}
\newcommand{\bvarphi}{\bar{\varphi}}
\newcommand{\bg}{\bar{g}}
\newcommand{\bV}{\bar{V}}
\newcommand{\ghost}{^{\phantom{a}}}

\usepackage{appendix}

\usepackage[retainorgcmds]{IEEEtrantools}

\usepackage{environ}

\NewEnviron{eqsplit}{
\begin{equation}
\begin{split}
	\BODY
\end{split}
\end{equation}
}

\NewEnviron{eqsplit*}{
  \begin{equation*}
  \begin{split}
  \BODY
  \end{split}
  \end{equation*}
}

\begin{document}

\begin{titlepage}
\begin{center}
\hfill NIKHEF 2014-025

\vskip 1.5cm
{\LARGE \bf Equivalence of the Einstein and Jordan frames}
\vskip 1cm

{\bf Marieke Postma$^{\dagger}$ \& Marco Volponi$^*$}
\vskip 25pt
{\em
\hskip -.1truecm  Nikhef, \\Science Park 105, \\1098 XG Amsterdam,\\ The
Netherlands \vskip 5pt }


\end{center}

\vskip 2cm

\begin{abstract}
  \noindent No experiment can measure an absolute scale: every
  dimensionfull quantity has to be compared to some fixed unit scale in
  order to be measured, and thus only dimensionless quantities are
  really physical.
	
  The Einstein and Jordan frame are related by a conformal
  transformation of the metric, which amounts to rescaling all
  length scales. Since the absolute scale cannot be measured, both
  frames describe the same physics, and are equivalent. In this
  article we make this explicit by rewriting the action in terms of
  dimensionless variables, which are invariant under a conformal
  transformation. For definiteness, we concentrate on the action of
  Higgs inflation, but the results can easily be generalized. In
  addition, we show that the action for f(R)-gravity, which includes
  Starobinsky inflation, can be written in a frame-independent form.

\end{abstract}

\vfill

{email: $\dagger$ {\tt mpostma@nikhef.nl} , $*$ {\tt mvolponi@nikhef.nl}} \\

\end{titlepage}



\section{Introduction}

In recent years there has been renewed interest in inflation models
with a large non-minimal coupling to gravity, of which Higgs inflation
is the prime example \cite{fakir,salopek,bezrukov1}. Although the
predictions of these models fall right in the sweet spot of the Planck
data \cite{Planck}, they can all go in the dust bin if the
polarization signal measured by BICEP is of cosmological origin
\cite{Bicep1,Bicep2}. Even in this case, a (much smaller) non-minimal
coupling is still allowed \cite{lowxi1,lowxi2,lowxi3}, and it is thus
important to understand its implications.

The non-minimal coupling to gravity entails a  coupling
between the Ricci scalar and the inflaton field, which mixes the
metric and scalar degrees of freedom. This also implies that the
effective Planck mass during inflation is field-dependent, and thus
time-dependent, which in turn hampers a physical interpretation of the
equations in the Jordan frame. For example, when defining the
expansion rate of the universe, one has to take into account that not
only the scale factor is time-dependent, but also the measurement
unit; when defining an energy-momentum tensor, one has to take
into account that gravitational and field energies are mixed, and so on.

The non-minimal coupling can be removed, and the gravity Lagrangian
brought in canonical form, by performing a conformal transformation of
the metric. Since now gravity is standard and the Planck mass a
constant, the Einstein frame equations are easy to interpret --- all
the usual textbook intuition applies --- but complicated. The scalar
field kinetic terms are non-canonical and the potential is
non-polynomial.

Calculations can be done in either frame. The conformal transformation
can just be seen as a field redefinition which does not affect the
physics.  It has been shown that both the classical action and the
one-loop corrections are frame-independent \cite{George1}, as well as
the curvature perturbation \cite{gong,chiba,kubota,janW1,janW2}.
Nevertheless, there is still some confusion in the literature and
conflicting claims exists
\cite{loop1,loop2,loop3,loop4,odintsov1,odintsov2,white1,white2}.  
Our results are in line with earlier works \cite{catena,hertzberg,quiros1,quiros2}.

In this paper we will show explicitly that the Jordan and Einstein
frame are equivalent, by rewriting the action in terms of
dimensionless fields and parameters. A conformal transformation of the
metric rescales all length scales, or equivalently all mass scales, in
the theory. It is important to note that this does not affect physical
quantities, which are dimensionless; no experiment can measure an
absolut scale. Hence, if we rewrite the action in terms of physical,
dimensionless fields, it is automatically invariant under a conformal
transformation: the action obtained describes all frames related by a
conformal transformation at once, and thus all the results derived
from it apply equally to the Jordan and to the Einstein frame. 

This paper is organised as follows: we start in section
\ref{s:equivalence} with a brief review of the action of Higgs
inflation in the Einstein and in the Jordan frame, and the conformal
transformation which relates them. We then explain our approach to
rewriting the Lagrangian in terms of dimensionless variables, applying
it to the simple setting of just the classical background action. This
is subsequently generalized to the full action, with a generic
spacetime metric, more than one field coupled to gravity, and
arbitrary kinetic terms. We shortly remark on quantization.  In
section \ref{s:starobinsky} we show that our results equally apply to
$f(R)$ gravity, and in particular Starobinsky inflation. Finally, in
section \ref{s:zeta} we give an example and discuss how curvature
perturbation $\zeta$ can be expressed in dimensionless, Jordan frame
and Einstein frame quantities. We end with some concluding remarks.

\section{Equivalence of the action}
\label{s:equivalence}

The action for Higgs inflation in the Jordan frame is \cite{fakir,salopek,bezrukov1}
\be
S = \int \dd^4 x \sqrt{-g_J\ghost} \[ \frac12 M^2 \Omega^2 R[g_J\ghost] - \frac12 \gamma\ghost_{Jab}
g_J^{\mu\nu} \partial_\mu \phi_J^a \partial_\nu \phi_J^b  - V_J\ghost(\phi_J\ghost)\] \nn \\
\label{S_J}
\ee
where $\phi_J^a$ are real scalar fields, $\xi$ is the non-minimal coupling that mixes the metric and scalar
degrees of freedom, and
\be \Omega^2 = 1+ \frac{\xi \phi_J^a \phi_{Ja}\ghost}{M^2} .
\label{xi}
\ee 
Although our focus is on the conformal factor \eref{xi} motivated by
Higgs inflation, our methods can be generalized to generic conformal
factors \cite{brax}.  The gravitational action can be brought in canonical form
via a conformal transformation of the metric
\be
g\ghost_{E\mu\nu} = \Omega^2 g\ghost_{J\mu\nu} 
\label{trafo}
\ee
This yields the action in the Einstein frame \cite{kaiser}
\be
S = \int \dd^4 x \sqrt{-g_E\ghost} \[ \frac12 M^2 R[g_E\ghost] - \frac12
g_E^{\mu\nu} \gamma\ghost_{Eab}\partial_\mu \phi_J^a \partial_\nu \phi_J^b  - V\ghost_E( \phi\ghost_J ) \]
\label{S_E}
\ee
with field-space metric
\be
\gamma\ghost_{E \, ab} = \frac1{\Omega^2}\[ \gamma_{Jab}\ghost +3 \partial_{\phi_J^a} \ln
\Omega^2 \partial_{\phi_J^b} \ln \Omega^2\].
\label{gammaE}
\ee
The Einstein frame potential is
\be
V_E = \frac{V_J}{\Omega^4}.
\label{VE}
\ee
Note that the fields $\phi_J^a$ do not change going from one frame to
the other. We added the subscript $J$ to denote the fields originally
defined in the Jordan frame; the use of this will become clear soon.

From the action we can read off the (reduced) Planck mass in the Jordan and
Einstein frame respectively:
\be
m_J = M \Omega, \qquad m_E = M.
\label{mplanck}
\ee
Although the metric changes under a conformal transformation,
distances measured in Planck units are invariant. Indeed,  the
line-element written in Planck units is invariant:
\be
m_J^2 \dd s_J^2 = m_J^2 g\ghost_{J\mu \nu} \dd x^\mu \dd x^{\nu}
=  m_E^2 g_{E\mu \nu}\ghost \dd x^\mu \dd x^{\nu}
= m_E^2 \dd s_E^2
\label{ds}
\ee
where we used (\ref{trafo},~\ref{mplanck}).

\subsection{Dimensionless action --- background}

In this subsection we rewrite the classical background action in terms
of dimensionless quantities which transform trivially under a
conformal transformation: this shows clearly our approach in a simple
setting. Then, in the next subsection, we will generalize our results to the
full action.

For simplicity, we take the field space metric in the Jordan frame to
be canonical $\gamma_{Jij}\ghost = \delta_{ij}$, and specialize to the case
of a single, homogeneous background field $\phi^a_J(x) = \phi_J(t)\ghost$. The
metric is of the FRW form, and \eref{ds} can be written
\be
m_J^2 \dd s_J^2 = m_J^2 [-N_J^2 \dd t^2 + a_J^2 \dd x^2]  =
m_E^2 [-N_E^2 \dd t^2 + a_E^2 \dd x^2]  = m_E^2 \dd s_E^2,
\ee
with $N_J,a_J$ and $N_E,a_E$ the lapse function and scale factor in
the Jordan and Einstein frame respectively.  We choose the coordinates
to be dimensionless and take $N_i,a_i$ to have dimensions of inverse
mass.\footnote{We could just as well have used the more standard
  convention with $a,N$ dimensionless and the coordinates dimensions
  of inverse mass: but in that case attention should be paid in defining the Hubble
  parameter, which should be taken as $H =
  \partial_t \ln \Delta x $ (i.e. the rate of change of a physical coordinates
  distance rather than of the unphysical scale factor $H = \partial_t
  \ln a$).}
We define the dimensionless metric functions, denoted by an overbar, via
\be
\bar N_i = m_i N_i, \qquad \bar a_i = m_i a_i, \qquad i = J,E
\ee
where $m_i$ is the frame-dependent Plank mass \eref{mplanck}.  The
dimensionless quantities transform trivially under a conformal, e.g.
$\bar N_J = \bar N_E$. To make this explicit we drop the subscript index on
the barred quantities and simply write $\bar N$, etc.  All barred
quantities defined below transform trivially; they correspond to the
respective quantities expressed in Planck units.

Now let us rewrite the background action in terms of  physical dimensionless
quantities. We start from the Einstein frame action \eref{S_E}, which
can be expressed
\be
S= \int \dd^4 x \sqrt{-g_E\ghost} m_E^4  \[ \frac12  \frac{R[g_E\ghost]}{m_E^2} -
\frac12 \gamma_E\ghost \frac{  {\dot{\phi}_J}^2}{N_E^2 m_E^4}  - \frac{V_J\ghost}{\Omega^4 m_E^4}\]
\label{Sdim1}
\ee
where $\gamma_E\ghost \equiv \gamma_{E \,\phi\phi}\ghost$, with the metric given in
\eref{gammaE}, and the dot denotes the time-derivative 
$\dot{\phi}_J\ghost =  \partial_t \phi_J\ghost$.

All the separate terms in the action \eref{Sdim1} and also the
measure are written as dimensionless combinations.  We are now going
to rewrite these terms in a form that makes explicit that
they are all separately invariant under a conformal transformation. 
Consider first the measure: we define the invariant combination
\be
\sqrt{-\bar g} = \sqrt{-g_i}m_i^4
,\qquad i =J,E.
\label{gbar}
\ee
It can be checked explicitly that it is invariant under a conformal
transformation $\sqrt{-\bar g} =\sqrt{-g_E\ghost} m_E^4 = N_E\ghost a_E^3 m_E^4 =
N_J\ghost a_J^3 m_J^4 = \sqrt{- g_J\ghost} m_J^4$.  The dimensionless potential
can be defined likewise
\be
\bar V= \frac{V_i}{m_i^4}
\label{Vbar}
\ee
where we used \eref{VE}.  Note that if $V_J\ghost = \lambda \phi_J^4$, this implies the scaling 
\be
\phi_E\ghost = \frac{\phi_J\ghost}{\Omega}.
\label{phibar}
\ee
This is in general the case: dimensionfull variables scale with a
factor $\Omega$ under a conformal transformation. As a consequence,
physical observables which are dimensionless ratios remain invariant.
Care should be taken though, when defining dimensionless quantities
involving time-derivatives, such as the Hubble constant. The reason is
that in the Jordan frame not only the quantity itself is
time-dependent, but also the measurement stick (e.g. when expressed in
Planck units, it is important to take into account that the Jordan
frame Planck mass \eref{mplanck} is time-dependent itself).  On the
background the Ricci scalar is $R_i = 6 (2H_i^2 + H'_i)$ with
$i=J,E$. To write this in physical quantities we define the
dimensionless Hubble constant
\be
\bar H =\frac{\bar a'}{\bar a}
=\frac{1}{\bar a}\frac{\partial_t \bar a}{\bar N} =
\frac{1}{a_i m_i} \frac{1}{m_i N_i }\partial_t (a_i m_i) 
\ee
with $i=E,J$ for Einstein and Jordan frame quantities respectively,
and the prime derivative is defined as
$\bar{\phi}'_i=\frac{1}{\bar{N}_i}\dot{\bar{\phi}}_i$. This
dimensionless Hubble constant transforms trivially under a conformal
transformation. Likewise we define
\be
 {\bar H}' = \frac{1}{\bar N} \partial_t \bar H
\ee
so we can write the dimensionless Ricci scalar (on the background) as
\be
\bar R = 6 (2\bar H^2 +  \bar H')
\label{Rbar}
\ee

To formulate the kinetic term in invariant form it is convenient to
first express the Einstein frame quantities in terms of the rescaled
field $\phi_E\ghost$ using \eref{phibar}. Then, in a next step, it is
straightforward to introduce the invariant and dimensionless field,
defined in the usual way
\be
\bar \phi = \frac{\phi_i}{m_i}, \qquad, \bar \phi' = \frac1{\bar
  N} \partial_t \bar \phi
\label{barphi}
\ee 
Before going further, let us reformulate $\Omega^2$ in terms of
the Einstein frame field:\footnote{
The dimensionless field is bounded from above
\be
 \xi \bar \phi^2 =  \frac{\xi
  \phi_J^2}{(\mpl^2 + \xi \phi_J^2)}< 1
\ee
as the denominator is always larger than the numerator.  It follows
that $\Omega^2$ in \eref{OmegaBG} is always positive
definite. \label{note:singular}}
\be
\Omega^2 = 1 +\frac{\xi \phi_J^2}{m_E^2}  
= 1 +\xi \bar \phi^2\Omega^2 
\qquad \Rightarrow \qquad \Omega^2  = \frac{1}{1-\xi \phi_E^2/m_E^2}
 = \frac{1}{1-\xi \bphi^2}.
\label{OmegaBG}
\ee
The derivatives of the Einstein and the Jordan frame fields $\phi_J\ghost
=\Omega \phi_E\ghost$ are related via
\begin{eqsplit}
\frac{1}{N_E}\dot{\phi}_J\ghost & = {\Omega}\( \phi_E'+ \phi_E\ghost \frac{\Omega'}{\Omega}\)
= \frac{\Omega}{1 - \xi \phi_E^2/m_E^2} \frac{1}{N_E}\dot{\phi}_E\ghost  \\
&=\Omega^3 \frac{1}{N_E}\dot{\phi}_E\ghost .
\label{phiprime}
\end{eqsplit}
The dimensionless expression for the kinetic terms in \eref{Sdim1} can
now be rewritten 
\begin{eqsplit}
\frac12 \frac{ \gamma_E\ghost \dot{\phi}_J^2}{N_E^2 m_E^4}  &= 
\frac1{2\Omega^2 N_E^2 m_E^4} \(1+ 6 
m_E^2 (\partial_{\phi_J\ghost} \Omega)^2\)
\dot{\phi}_J^2
=\frac{\Omega^4}{2} \(1+ 6 \frac{\xi^2 \phi_E^2}{ m_E^2}\)
\frac{\dot{\phi}_E^2}{N_E^2 m_E^4} \\
& =\frac{\Omega^4}{2} \(1+ 6{\xi^2 \bar \phi}\)
{\bar{\phi'}}^{2} \equiv \frac12\bar S(\bar \phi) {\bar{\phi'}}^2 .
\label{kinbar}
\end{eqsplit}
In the second expression we used the explicit form of the Einstein
frame metric \eref{gammaE}, and in the third we used the relations
between the fields in the two frames
(\ref{phibar},\ref{phiprime}). Finally, in the last two expressions we
introduced the frame-invariant fields \eref{barphi}.
$\bar S(\bar \phi)$ is the dimensionless and frame-invariant field-space metric, which
is a function of the dimensionless field $\bar \phi$.\footnote{$\bar S$ is
  not directly related to either $\gamma_E\ghost$ (because the fields in
  \eref{S_E} are still the jordan frame fields) or $\gamma_J\ghost$ (when
  writing out the $\bar S$ and $\bar R$ terms in Jordan frame
  quantities, both contribute to the Jordan field kinetic terms $\gamma_J\ghost$).}

Now we have all the expressions
(\ref{gbar},\ref{Vbar},\ref{Rbar},\ref{kinbar}) needed to write the action in terms of
dimensionless quantities, which reads
\be
S= \int \dd^4 x \sqrt{-\bar g}   \[ \frac12  \bar R(\bar a,\bar N) -
\frac12 \bar S (\bar \phi) {\bar{\phi'}}^2  - \bar V(\bar \phi)\].
\label{S_background}
\ee
Making a conformal transformation leaves the action invariant, therefore the latter describes
all frames related by a conformal transformation.  In fact all
relevant equations and expressions can be derived
from this action. If at some point it is desired to express them into
frame dependent quantities, it can easily be done by substituting the
explicit definitions of the barred quantities.  This way it can be
checked that \eref{S_background} indeed returns to the Jordan frame
action, when the Planck mass and variables proper to that frame are substituted.

We have expressed all quantities in Planck units. This a very
convenient choice for calculations in cosmology, and moreover, the
results can readily be compare with experiments. Of course, the choice
of units is not unique. The reason \eref{S_background} takes the form
of the Einstein frame action is precisely because in that frame the Planck
mass (our reference) mass is constant.

\subsection{Dimensionless action --- full action}

In the previous section we have shown our idea at work in a very 
simple but significant example: now we want to extend the results 
to the full action (not just the background) and
allow for several non-minimally coupled scalar fields.  

The metric and scalar fields are thus taken both time and space
dependent; we keep the Jordan frame field metric $\gamma_J\ghost$ and
potential $V_J$.


The approach is the same as before: start with the Einstein frame
action \eref{S_E}, and write it in terms of the Einstein frame fields
$\phi_J^a = \Omega \phi^a_E$; in the next step, go to dimensionless
and frame-invariant variables by dividing everything with the proper
powers of the Planck mass.  Care should be taken for quantities that
involve derivatives: the derivatives should always act on dimensionless
and frame-invariant quantities themselves; this takes properly into account
that the Planck mass is space-time dependent in the Jordan frame.

The only non-trivial step is to relate the derivatives of the Jordan
and Einstein frame fields, and rewrite the kinetic terms.  It is
convenient to start expressing $\Omega$ in terms of Einstein frame
fields.
\begin{eqsplit}
	&\Omega^2 = 1+ \frac{\xi}{m_E^2} \phi^a_J \phi_{Ja}\ghost = 1+ \frac{\xi}{m_E^2} \phi^a_E \phi_{Ea}\ghost \Omega^2 \\
	\Rightarrow \quad  &\Omega^2 = \left( 1-\frac{\xi}{m_E^2}
          \phi^a_E \phi_{Ea}\ghost \right)^{-1}= \left( 1-\xi
          \bar\phi^a \bar \phi_{a} \ghost\right)^{-1}
\end{eqsplit}
In the 2nd line $\xi \bar \phi^a \bar \phi_{a}\ghost < 1$ always,
see footnote \ref{note:singular}.  Now we can proceed
\begin{equation}
		\begin{split}
			\nablamu \phi^a_J &= \phi^a_E \nablamu \Omega
                        + \Omega \nablamu \phi^a_E
 =  \Omega \left( \delta^a_c +
                          \frac{\xi}{m_E^2}\Omega^2 \phi^a_E
                        \phi_{Ec}\ghost  \right) \nablamu \phi^c_{E} 
 \\
			&\equiv \M^a_c \nablamu \phi^c_E . 
		\end{split}
\end{equation}
Finally, the field-space metric tensor for the Einstein frame fields can be calculated:
\begin{align}
		S_{ab} &= \gamma_{Ecd}\ghost \M_a^c \M^d_b = \\
		&= \Omega^{-2} \left( \gamma_{Jcd}\ghost + 6 \frac{\xi}{m_E^2} \phi_{Ec}\ghost \phi_{Ed}\ghost \right) \Omega^2 \left(\delta_a^c + \Omega^2 \frac{\xi}{m_E^2} \phi^c_E \phi_{Ea}\ghost \right) \left(\delta_b^d + \Omega^2 \frac{\xi}{m_E^2} \phi^d_E \phi_{Eb}\ghost \right)  \\
		&=\gamma_{Jab}\ghost +\frac{\xi}{m_E^2} \Omega^2
                \(\gamma_{Jad}\ghost \phi_E^d
                \phi_{Eb}\ghost+\gamma_{Jbd}\ghost \phi_E^d
                \phi_{Ea}\ghost\) + \Omega^4 \phi_{Ea}\ghost
                \phi_{Eb}\ghost \(\gamma_{Jcd}\ghost \phi_E^c \phi_E^d
                \frac{\xi^2}{m_E^4}+ 6\frac{\xi}{m_E^2}\) \nn
\end{align}
As it should, in the single field limit $\phi_E^a =\phi_E\ghost$ and for a
trivial field metric $\gamma_{J  ab}\ghost =\delta_{ab}$, the expression reduces,
after some simplifications, to the field space metric found in the
previous section \eref{kinbar}.

Now it is clear how to pass to the dimensionless action and
frame invariant action.  We define the frame-independent fields
\be
\bar \phi^a = \frac{\phi_{ia}\ghost}{m_i}, \qquad
\bar{g}_{\mu\nu}\ghost = g_{i\mu\nu}\ghost m_i^2, 
\ee
with $i=J,E$. All other quantities are constructed from these:
\begin{IEEEeqnarray}{LCL}
	\bar{V} &=& \frac{V_i}{m_i^4}; \\
	\bar{\Gamma}^{\sigma}_{\alpha\beta} &=& \frac12
        \bar{g}^{\sigma\rho} \left( \partial_{\alpha}
          \bar{g}_{\beta\sigma}  + \partial_{\beta}\bar{g}_{\alpha\sigma}
          - \partial_{\sigma}\bar{g}_{\alpha\beta} \right) 
\label{V_bar}  \\
	\bar{R} &=& \bar{g}^{\alpha\beta} \left( \partial_{\sigma} \bar{\Gamma}^{\sigma}_{\alpha \beta} - \partial_{\beta} \bar{\Gamma}^{\sigma}_{\alpha \sigma} + \bar{\Gamma}^{\sigma}_{\alpha \beta} \bar{\Gamma}^{\rho}_{\sigma\rho} - \bar{\Gamma}^{\rho}_{\alpha\sigma} \bar{\Gamma}^{\sigma}_{\beta\rho} \right)  \\
	\bar{S}_{ab} &=&\gamma_{Jab}\ghost + \xi \Omega^2 \(\gamma\ghost_{Jad}
        \bar\phi^d \bar \phi_{b}\ghost
+\gamma\ghost_{Jbd} \bar\phi^d \bar\phi_{a}\ghost\) + \Omega^4 \bar\phi_{a}\ghost \bar \phi_{b}\ghost \(
\gamma\ghost_{Jcd} \bar \phi^c \bar \phi^d \xi^2+ 6\xi\).
\label{S_bar}
\end{IEEEeqnarray}
Choosing $i=E$ and substituting in the Einstein frame action
\eref{S_E} finally gives the action in explicitly frame-independent
and dimensionless form:
\begin{equation}
	S=\int d^4 x \sqrt{-\bar{g}} \left( \frac12 \bar{R} - \frac12
          \bar{S}_{ab}\bar{g}_{\mu\nu} \nablamu \bar \phi^a \nablanu \bar\phi^b - \bar{V} \right) .
\label{S_final}
\end{equation}
This is our main result: the action in written in a frame-invariant
form, so all equations derived from it apply equally to all
actions related by a conformal transformation; moreover, the results
can readily be related to experiments, which only measure
dimensionless quantities. In practice, we can simply take the usual
Einstein frame results, set the Planck mass to unity $m_E =1$, and
put a bar on all quantities: this gives the frame-invariant equations.

\subsection{Quantization}

The discussion in the previous section was fully classical: we showed
that the classical action can be written in manifestly frame-invariant
form. But someone might still be worried that quantization introduces a frame
dependence: however, it is clear that if the quantization prescription
is formulated in terms of the frame-independent barred quantities, no
such issues arise.  We can thus use the standard quantization
procedures, applied to the action \eref{S_final}.

\section{f(R) gravity}
\label{s:starobinsky}

In this section we show that theories of $f(R)$ gravity, or
equivalently scalar-tensor gravity with the Brans-Dicke parameter
$\omega_{BD} =0$, can also be written in a frame-independent way.  Key
here is to realize that this class of theories can be rewritten as a
as a scalar theory with a non-minimal coupling to gravity
\cite{BD1,BD2,BD3,BD4,ketov}; then the frame-invariant approach of the
previous subsection can be applied.

Consider the action of $f(R)$-gravity
\be
S = \frac{M^2}{2} \int \dd^4 x \sqrt{-g_J\ghost} f(R)
\label{S_f}
\ee
whose function $f(R)$ begins with the Einstein-Hilbert
term. Starobinsky inflation is a specific example with $f(R) = R +
\alpha R^2$ \cite{starobinsky,vilenkin}. Introducing an auxiliary
scalar field $A_J$ the action can be rewritten \cite{ketov}
\be
S = \frac{M^2}{2} \int \dd^4 x \sqrt{-g_J\ghost} \[ A_J R - V_J(A_J) \]
\label{S_f2}
\ee
Applying the equations of motion for the scalar $R = \partial_A V$,
substituting in the action, one retrieves the original $f(R)$ action
\eref{S_f}, provided that $f$ and $V$ are related by a Legendre
transformation
\be
f(R)  = R A_J - V_J (A_J)
\ee

Now the  action \eref{S_f2} is exactly of the form of the Jordan frame
action \eref{S_J} for a single field, if we identify 
\be
\gamma_{Jab}\ghost = 0, \qquad A_J = \Omega^2 (\phi_J\ghost)
\ee
One can make a conformal transformation \eref{trafo} to go to the
Einstein frame. Using the results of the previous section (\ref{V_bar}
- \ref{S_bar}) the action can be written in explicitly dimensionless
and frame-invariant form.

\section{An example: equivalence of the curvature and isocurvature perturbations}
\label{s:zeta}

In the literature there are calculations of the curvature and
isocurvature perturbations in both the Einstein and Jordan frame. It
was shown that the curvature perturbation is frame-independent
\cite{gong,chiba,kubota,janW1,janW2} in the absence of curvature
perturbations. Hoewever, frame-dependent definitions were introduced
in the presence of isocurvature perturbations obscuring the
equivalence of the two frames \cite{white1,white2}. Applying the
results of the previous section, we can express the perturbations
spectrum in terms of the barred fields, which manifestly shows their
frame-independence.

The frame-independent perturbations can be rewritten in either
Einstein or Jordan frame quantities, using the definitions of the
barred quantities; the complicated relations between the two frames
shows how easy it to make mistakes when comparing results in different
frames, when they are not written in physical, dimensionless quantities.

To write the perturbation spectrum we have to perturb the field and metric to
first order
\begin{align}
	\bPhi^a (t,x) &= \bphi^a(t) + \bar{\varphi}^a(t,x), \qquad 
&		\bar{g}_{00} &= -\bar{N}^2 (1+2\bar{n}); \nn \\
&&		\bar{g}_{i0} &= \bar{g}_{0i} = 2\bar{a}\bar{N}\bar{n}_i;\nn \\
&&		\bar{g}_{ij} &= \bar{a}^2 (1-2 \bar{\psi}) \delta_{ij}
+ \bar{F}_{ij}.
\label{metric}
\end{align}
With this, we can express the gauge invariant scalar curvature perturbation as 
\begin{equation}
	\zeta = - \bar{\psi} - \bar H \frac{\delta \bar{\rho}}{\dot{\bar{\rho}}} .
\end{equation}
Note that we have not put a bar over $\zeta$ because it is both invariant and dimensionless.
The energy-density appearing in this equation is defined in the usual way
\be
\bar{\rho} = -\bg^{0\nu} \bar{T}_{\nu0} 
= \bg^{0\nu} \frac{2}{\sqrt{\bg}} \frac{\delta \bar{S}_M}{\delta
  \bg^{\mu\nu}} \delta_0^{\mu} 
= \bar{S}_{ab} \left[ \frac12 \bg^{\alpha\beta} \pa_{\alpha} \bPhi^a \pa_{\nu} \bPhi^b - \bg^{0\nu} \pa_{\nu} \bPhi^a \pa_{0} \bPhi^b \right] + \bV . 
\ee
In the presence of isocurvature perturbations the curvature
perturbation is non-conserved 
\be
\zeta' = -\frac{\bar H}{\bar \rho +p} \delta \bar p_{\rm nad},
\ee
with as before $\zeta' = \dot \zeta/\bar N$ the dimensionless
time-derivative, and $\bar p_{\rm nad}$ the non-adiabatic pressure
\be
\delta \bar p_{\rm nad}= \delta \bar p - \frac{\dot{\bar p}}{\dot{ \bar
  \rho}} \delta \bar \rho.
\ee 

\subsection{Invariance of $\zeta$} \label{zetainv}

In this subsection we write the curvature perturbation in Jordan and
Einstein frame variables, and show how these are related.

First we have to find the transformation between $\bar{\psi} =
\psi_E$\footnote{Clearly $\bar{\psi} = \psi_E$ as $\psi_E$ is
  dimensionless.} and $ \psi_J$.  We can express the invariant metric
in either Jordan or Einstein frame variables 
$\bar{g}_{\mu\nu} = M^2 \bg_{E\mu\nu} = M^2 \Omega^2 \bg_{J\mu\nu}
$. Expanding this relation to first order then gives
\begin{eqsplit}
 a_E^2 \left( 1-2 \psi_E \right) &= a_J^2 \Omega^2 (1-2 \psi_J) 
  = a_J^2 \left( 1 + \frac{\xi}{M^2}\phi_J^a\phi_{Ja}\ghost + 2 \frac{\xi}{M^2}\phi_J^a\varphi_{Ja}\ghost \right) (1-2 \psi_J)  \\
	&\simeq a_J^2 \Omega_{(0)}^2 \left( 1-2\psi_J + \frac{2 \frac{\xi}{M^2}\phi_J^a\varphi\ghost_{Ja}}{\Omega_{(0)}^2} \right)
\end{eqsplit}
up to second order in perturbation . Further, we defined $\Omega_{(0)}^2 = 1+\frac{\xi}{M^2}\phi_J^a\phi_{Ja}\ghost$. It follows that
\be
	\psi_E = \psi_J - \frac{1}{\Omega_{(0)}^2}
        \frac{\xi}{M^2}\phi_J^a\varphi_{Ja}\ghost
        = \psi_J - \frac{1}{2 \Omega_{(0)}^2} \pa_a \left(\Omega_{(0)}^2\right) \varphi_J^a . \\
\ee
Using the slow-roll approximation we can write
\begin{align}
	\dot{\bar{\rho} } 
        &\simeq \pa_t \left[ \frac{1}{2\bar{N}^2} 
          \bar{S}_{(0)ab} \dot{\bphi}^a\dot{\bphi}^b + \dot{\bV}_{(0)} \right] 
	\simeq \dot{\bV}^{(0)} 
	= 4 \lambda \bphi^2 \bphi^a \dot{\bphi}_a ; \nn \\
	\delta \bar{\rho} &\simeq \delta \bV 
	 = 4 \lambda \bphi^2 \bphi^a \bvarphi_a,
\end{align}
where in the last line we have approximated $\bar{\rho}$ with its
background value, using the expansion expressed in \eref{metric}; to
simplify notation we let $\phi^2 = \phi^b \phi^a \delta_{ab}$. Now we
relate the ratio $\delta \bar \rho/\dot{\bar \rho}$ in the two frames
\begin{eqsplit}
	\frac{\delta\rho_E}{\dot{\rho}_E} &= \frac{\delta\left(\lambda \left(\phi^2_J\right)^2\right) \Omega^2 -\lambda (\phi_J^2)^2 \delta \Omega^2  }{\lambda \Omega^2 \pa_t (\phi_J^2)^2 - \lambda (\phi^2)^2 \pa_t \Omega^2}  \\
	&= \frac{4 \lambda \left( 1 + \frac{\xi}{M^2}\phi_J^2 \right) \phi_J^2 \phi_J^a\varphi_{Ja}\ghost - 2 \lambda \left(\phi^2_J\right)^2 \frac{\xi}{M^2} \phi_J^2 \phi_J^a\varphi_{Ja}\ghost }{4 \lambda \left( 1 + \frac{\xi}{M^2}\phi_J^2 \right) \phi_J^2 \phi_J^a\dot{\phi}_{Ja}\ghost - 2 \lambda \left(\phi^2_J\right)^2 \frac{\xi}{M^2} \phi_J^2 \phi_J^a\dot{\phi}_{Ja}\ghost} \\
	&= \frac{2 \lambda \frac{1 + \Omega^2}{\Omega^2} \phi_J^2
          \phi_J^a\varphi_{Ja}\ghost }{2 \lambda \frac{1 +
            \Omega^2}{\Omega^2} \phi_J^2
          \phi_J^a\dot{\phi}_{Ja}\ghost} 
        = \frac{\phi_J^a\varphi_{Ja}\ghost}{\phi_J^a\dot{\phi}_{Ja}\ghost} = \frac{\delta\rho_J}{\dot{\rho}_J}.
\end{eqsplit}
So finally
\begin{eqsplit}
	\zeta&= -\psi_E - \frac{\dot a _E}{a_E}\frac{\delta\rho_E}{\dot{\rho}_E}  \\
	&= -\psi_J + \frac{1}{2 \Omega_{(0)}^2} \pa_a \left(\Omega_{(0)}^2\right) \varphi_J^a - \frac{1}{\Omega a_J}\left(\Omega \dot a _J + \dot{\Omega} a_J \right) \frac{\phi_J^a\varphi_{Ja}\ghost}{\phi_J^a\dot{\phi}_{Ja}\ghost} \\
	&= -\psi_J - \frac{\dot a _J}{a_J}\frac{\delta\rho_J}{\dot{\rho}_J} + \frac{1}{2 \Omega_{(0)}^2} \pa_a \left(\Omega_{(0)}^2\right) \varphi_J^a  - \frac{\dot{\Omega}}{\Omega} \frac{\phi_J^a\varphi_{Ja}\ghost}{\phi_J^a\dot{\phi}_{Ja}\ghost} \\
	&= \zeta .
\end{eqsplit}

\section{Conclusions}

Higgs inflation has attracted considerable interest over recent
years. The key ingredient of the model is a non-minimal coupling of
the Higgs field to the Ricci tensor. Unfortunately, this brings along with it the
issue of frames. The freedom to carry out the desired calculation in a
given frame without worrying about possible implications raised from
the choice of the frame itself is very important. With this in mind,
in this paper we have demonstrated the equivalence between the Jordan
and the Einstein frame, and more generally between every frame related
to these by a conformal transformation.

The equivalence of the various frames is not immediately obvious:
directly transforming quantities calculated in one frame to another
can (and have) lead to mismatching results. Seen from a physics
perspective, these frame-dependent results make no sense.  The key
point is that the conformal transformation that relates the Einstein
and Jordan frame rescales all length scales. Since the absolute scale
cannot be measured, both frames describe the same physics, and must be
equivalent.  It is thus important to realize that when applying a
conformal transformation, not only the spacetime changes, but also the
unit of measure is modified.  Therefore it is not surprising that when
comparing quantities between two frames without changing the measuring
reference accordingly, one obtains different results.

To avoid the unpleasant consequences of not keeping track of all scale
changes, we have introduced the concept of dimensionless variables: in
particular, we have chosen to express all dimensionfull quantities in
terms of Planck units. When transforming between frames, all mass
scales including the Planck mass scale in the same way; the
dimensionless ratios --- our dimensional variables --- remain however
the same. Rewriting the Lagrangian terms of these dimensionless
quantities provides a manifestly frame-invariant formulation of the
theory, which can be directly related to what is actually measured in
experiments. Moreover, formulating the action and all equations
derived from it in terms of dimensionless variables is very
convenient, because in any moment it is possible to choose a frame and
immediately convert the desired quantities into their frame-specific
counterparts by simply substituting the expressions for variables and
for the Plank mass proper of that frame. In the last section we have
given an example of this mechanism, showing how it works for the gauge
invariant curvature perturbation $\zeta$.

Our results are are applicable to $f(R)$ gravity, and in particular,
Starobinsky inflation. Introducing an auxiliary field these type of
actions can be written as a Jordan frame Lagrangian with a non-minimal
coupling to gravity. These can the in turn be expressed in our
dimensionless, physical quantities.
 

\section*{Acknowledgements}

The authors are supported by the Netherlands Foundation for
Fundamental Research of Matter (FOM) and the Netherlands
Organisation for Scientific Research (NWO). We thank
Damien George and Sander Mooij for useful discussions and a careful
reading of an earlier draft.

\end{document}